\documentclass[journal,onecolumn]{IEEEtran}
\usepackage{amsmath,amsfonts}
\usepackage{algorithmic}
\usepackage{algorithm}
\usepackage{array}
\usepackage[caption=false,font=normalsize,labelfont=sf,textfont=sf]{subfig}
\usepackage{textcomp}
\usepackage{stfloats}
\usepackage{url}
\usepackage{verbatim}
\usepackage{graphicx}
\usepackage{cite}
\usepackage{enumerate}
\usepackage{bm}
\usepackage{multirow}
\usepackage{color}
\usepackage{makecell}
\usepackage{CJK}
\usepackage{indentfirst}
\usepackage{cases}

\usepackage{xcolor}
\usepackage{xpatch}
\usepackage{footmisc}

\makeatletter
\def\changeBibColor#1{%	
\in@{#1}{kim2009design,richardson2001efficient,golmohammadi2021concatenated,b14,lu20163}% list of colored bib items	
\ifin@\color{black}\else\normalcolor\fi
}

\xpatchcmd\@bibitem
{\item}
{\changeBibColor{#1}\item}
{}{\fail}

\xpatchcmd\@lbibitem
{\item}
{\changeBibColor{#2}\item}
{}{\fail}
\makeatother



\makeatletter
\def\endthebibliography{%
\def\@noitemerr{\@latex@warning{Empty `thebibliography' environment}}%
\endlist
}
\makeatother

\newlength\myindent
\begin{document}

\title{\LARGE \color{black}Joint Design of {\color{black}Source-Channel} Codes with Linear
Source Encoding Complexity and Good Channel Thresholds Based on Double-Protograph LDPC Codes  \\
\thanks{
%%Part of the paper has been presented in ISTC 2021 \cite{lau2021joint}. \\
%%\indent 
The authors are with the %Future Wireless Networks and IoT Focusing Area,
Department of Electrical and Electronic Engineering,
The Hong Kong Polytechnic University, Hong Kong SAR, China.
(Emails: jia1206.zhan@connect.polyu.hk and francis-cm.lau@polyu.edu.hk.) 
\\ \indent The work described in this paper was partially supported by the RGC Research Impact Fund from the Hong Kong SAR, China (Project No. R5013-19).
}
}

\author{\IEEEauthorblockN{Jia Zhan and Francis C.~M. Lau \IEEEmembership{Fellow,~IEEE}} }
%\IEEEauthorblockA{\textit{}
%\author{IEEE Publication Technology,~\IEEEmembership{Staff,~IEEE,}
%        % <-this % stops a space
%\thanks{This paper was produced by the IEEE Publication Technology Group. They are in Piscataway, NJ.}% <-this % stops a space
%\thanks{Manuscript received April 19, 2021; revised August 16, 2021.}}


\maketitle

\begin{abstract}
We propose the use of a lower or upper triangular sub-base matrix to replace the identity
matrix in the source-check-channel-variable linking protomatrix of a 
double-protograph low-density parity-check joint-source-channel code (DP-LDPC JSCC). 
The elements along the diagonal of the proposed lower or upper triangular sub-base matrix
are assigned as ``1'' and the other non-zero elements can take any non-negative integral values. 
Compared with the traditional DP-LDPC JSCC designs, 
the new designs show a theoretical channel threshold improvement of up to $0.41$~dB
and a simulated source symbol error rate improvement of up to $0.5$~dB at an
error rate of $10^{-6}$.
\end{abstract}


\section{Introduction}
Designing source code and channel code separately is optimal when the code length is very long 
\cite{shannon1948mathematical}. For application scenarios requiring short to moderate code lengths, 
designing source code and channel code jointly can provide a higher {\color{black}coding} gain. 
The main idea of jointly designing source-channel code (JSCC) is to exploit the residual redundancy of the source in the tandem joint source-channel encoding/decoding algorithms so as to achieve coding gains \cite{hagenauer1995source}.
%It has been shown in \cite{hagenauer1995source} that considerable coding gains can be obtained by providing the prior probability of the source bits to the channel decoder. 
%%In the source-channel coding system, the source code and channel code can be designed jointly to optimize overall error performance. 

Low-density parity-check (LDPC) codes have been widely used in many communication systems 
%\cite{liva2006design,thangaraj2007} 
because of their capacity-approaching error correction capability \cite{mackay1996near}. 
%{\color{black}They have been widely used in many communication systems\cite{liva2006design,thangaraj2007}.}
In \cite{fre2010joint}, LDPC codes are used as both the source code and the 
channel code in a JSCC system, where the source encoder and channel encoder are concatenated in series, and a joint decoder with information exchange between the source decoder and the channel decoder is utilized. 
%Protograph-based LDPC (P-LDPC) codes \cite{divsalar2005protograph,divsalar2009capacity} are a subset of LDPC codes and {\color{black}their structure facilitates parallel encoding and decoding}. 
In \cite{he2012joint}, the LDPC codes in the JSCC system are replaced by protograph-based LDPC (P-LDPC) codes  
\cite{divsalar2009capacity} {to form} the double-protograph-based LDPC JSCC (DP-LDPC JSCC) system.

The protograph representation of a traditional DP-LDPC JSCC system is shown in Fig.~\ref{JSCC_DPLDPC_modified} (without the red connections). 
Variable nodes (VNs) and check nodes (CNs) are denoted by circles and squares, respectively. 
%The green lines connect CNs in the source P-LDPC code and VNs in the channel P-LDPC code in a one-node-to-one-node manner, reflecting the cascading relationship between the source encoder and the channel encoder. 
The traditional DP-LDPC JSCC can also be represented 
by a joint protomatrix $\textbf{B}_{\mathrm J_0}$, i.e.,  
$\textbf{B}_{\mathrm J_0}=
\begin{pmatrix} 
\makebox[-0.5cm]{} \textbf{B}_{\mathrm s} &  \textbf{B}_{\mathrm {sccv}} \\
\;\; \textbf{0}_{m_c\times n_s} & \makebox[-0.4cm]{}  \textbf{B}_{\mathrm c}
\end{pmatrix}$,
%\begin{equation}
%\textbf{B}_{\mathrm J_0}=
%\begin{pmatrix} 
%\makebox[-0.6cm]{} \textbf{B}_{\mathrm s} &  \textbf{B}_{\mathrm {sccv}} \\
%\;\; \textbf{0}_{m_c\times n_s} & \makebox[-0.5cm]{}  \textbf{B}_{\mathrm c}
%\end{pmatrix}
%\label{eq:B_J0}
%\end{equation}
where 
%$B_s \in \mathbb{B}^{m_s \times n_s}$ indicates the source protomatrix
$\textbf{B}_{\mathrm s}$ indicates the source protomatrix with size $m_s\times n_s$, 
$\textbf{B}_{\mathrm c}$ indicates the channel protomatrix with size $m_c\times n_c$, 
and $\textbf{B}_{\mathrm {sccv}}$ indicates the source-check-channel-variable (SCCV) linking protomatrix 
\cite{chen2020analysis}
with size $m_s\times n_c$. Moreover, $\textbf{B}_{\mathrm {sccv}}$ is given by 
\begin{equation}
\textbf{B}_{\mathrm {sccv}}=\begin{pmatrix} \textbf{0}_{m_s\times m_c} & \textbf{I}_{m_s}\end{pmatrix},
\label{eq:constraint}
\end{equation}
where 
$\textbf{0}_{m_s\times m_c}$ denotes a zero matrix with size $m_s\times m_c$, and $\textbf{I}_{m_s}$ indicates an identity matrix with size $m_s\times m_s$.

To evaluate the channel thresholds and source thresholds of the DP-LDPC JSCCs\footnote{\color{black} 
Strictly speaking, the code structure is not fully a JSCC
because the source P-LDPC code and the channel P-LDPC code are concatenated through 
the SCCV linking protomatrix $\textbf{B}_{\mathrm {sccv}}$ and are encoded separately.  
%However the %source residual redundancy is exploited by the channel decoder,
%%and updated extrinsic information is exchanged between the ``source decoder'' and the ``channel decoder'' iteratively. 
%%The 
%above 
But it has been loosely categorized as the ``joint source-channel coding'' problem because the source code 
and channel code are jointly decoded \cite{fre2010joint}.
Here, we follow the loose definition of JSCC and continue to name such codes DP-LDPC JSCCs.
%``double-protograph-based LDPC joint-source-channel codes (DP-LDPC JSCCs).'' 
}, a joint protograph extrinsic information transfer (JPEXIT) algorithm and a source protograph 
extrinsic information transfer 
(SPEXIT) algorithm are proposed %\cite{chen2016performance,chen2015matching}.
in \cite{chen2016performance} and \cite{chen2015matching}, respectively. 
Moreover, 
various optimizations on the DP-LDPC JSCC system have been performed \cite{chen2016performance,chen2015matching,chen2018joint,chen2018design,liu2020joint,Chen2019}, e.g.,  
%{\color{black}
%In  \cite{b14}, an unequal power allocation technique is applied to the DP-LDPC JSCC system to improve its error performance.} Another method is to redesign the source protograph and/or channel protograph in the DP-LDPC JSCC system so that better error performance can be attained. 
%For example, 
the channel protograph $\textbf{B}_{\mathrm c}$ is redesigned based on the JPEXIT algorithm  \cite{chen2016performance};
the source protograph $\textbf{B}_{\mathrm s}$ is redesigned based on the SPEXIT algorithm  \cite{chen2018design};
%In \cite{chen2018joint}, the source and channel protograph pairs are redesigned to achieve good error-correction capability. In \cite{Chen2019}, the optimal distribution of degree-2 VNs in both the source and channel protographs is investigated.
% In \cite{liu2020joint}, 
and the joint protomatrix $\textbf{B}_{\mathrm J_0}$ is redesigned by considering both the source threshold and channel threshold \cite{liu2020joint}. 
%{\color{black}In \cite{golmohammadi2021concatenated}, double spatially-coupled LDPC codes are proposed to replace DP-LDPC codes in the JSCC system. Results show that
%with a sliding window decoding algorithm, low latency and good error performance
%can be obtained.} 
%
All the above DP-LDPC JSCC systems are required to satisfy the
common constraint shown in \eqref{eq:constraint}. 
{\color{black} Such a constraint allows easy encoding but 
largely {\color{black}restricts} the design flexibility. Hence the channel thresholds may not be optimized.  
{\color{black}In \cite{lau2021joint}, a class of JSCCs based on a single protograph is proposed.
The proposed protograph-based JSCCs (P-JSCCs)
%(i.e., P-JSCCs called AR3A-JSCC 
%code and AR4JA-JSCC code) 
are shown to outperform the 
optimized DP-LDPC JSCCs in \cite{liu2020joint}. Due to the removal of structural constraints such as that in \eqref{eq:constraint},  
the encoding complexity of P-JSCCs is higher than that of DP-LDPC JSCCs.}

% When $\textbf{B}_{\mathrm {sccv}}$ has the structure shown in \eqref{eq:constraint}, source compressed symbols can be obtained by using some simple modulo-2 addition operations. These source-compressed symbols are the inputs for the channel encoder. When $\textbf{B}_{\mathrm {sccv}}$ does not have the constraint of \eqref{eq:constraint}, we need to use the traditional channel encoding method to obtain channel codewords based on the joint base matrix. The parity-check matrix generated based on the single protograph is first transformed to a systematic form by using Gaussian elimination. The complexity of encoding is $\mathcal{O}(N^2)$ where $N$ is the length of source symbols and channel codewords.

{\color{black}In the traditional LDPC channel encoding, linear and fast encoding can be performed when the parity portion of the parity-check matrix possesses a lower or upper triangular structure \cite{kim2009design,richardson2001efficient}. In this paper, 
%we aim to increase the code design flexibility of the traditional DP-JSCC JSCCs to obtain lower channel thresholds and preserve the linear and fast source encoding property, 
we propose a new class of DP-LDPC JSCC which replaces the identity matrix $\textbf{I}_{m_s}$ in the SCCV linking protomatrix $\textbf{B}_{\mathrm {sccv}}$ with a lower or upper triangular sub-base matrix. 
With the increased code design
flexibility, we can design DP-LDPC JSCCs with better channel
thresholds while preserving the linear and fast source encoding
property. 
%We also evaluate the proposed DP-LDPC JSCCs' encoder/decoder
%complexity and latency.
%To illustrate the feasibility of our scheme, we reconstructed some DP-LDPC JSCCs based on source-channel pairs optimized in \cite{liu2020joint}. We also evaluate the increase in encoder/decoder complexity and latency.
}
%We show that the complexity of the encoding/decoding process of the new DP-LDPC JSCC is marginally increased
%and the error performance is improved by up to $0.5$~dB at a symbol
%error rate of $10^{-6}$.

%All the aforementioned DP-LDPC JSCC systems are required to satisfy the
%common constraint shown in \eqref{eq:constraint}. 
%{\color{black} Such a constraint allows easy encoding but 
%	largely {\color{black}restricts} the design flexibility. Hence the channel thresholds may not be optimized. 

%Some cases are given to show the feasibility and effectiveness of this scheme. 
We organize this work as follows. 
Section~\ref{sect:DP-LDPC JSCC} shows the details of the new class of DP-LDPC JSCC and analyzes its encoder/decoder complexity and latency. 
In Section~\ref{sect:results}, we construct new DP-LDPC JSCCs and compare their performance with traditional ones {\color{black}as well as the P-JSCCs in \cite{lau2021joint}}. 
%We also evaluate their channel thresholds and simulate their error rate; and compare them with those of the original DP-LDPC JSCCs. 
Finally, we give some concluding remarks %and some future directions 
in Section~\ref{sect:conclusion}.

\begin{figure}[t]
\centering
{\includegraphics[width=3.0in]{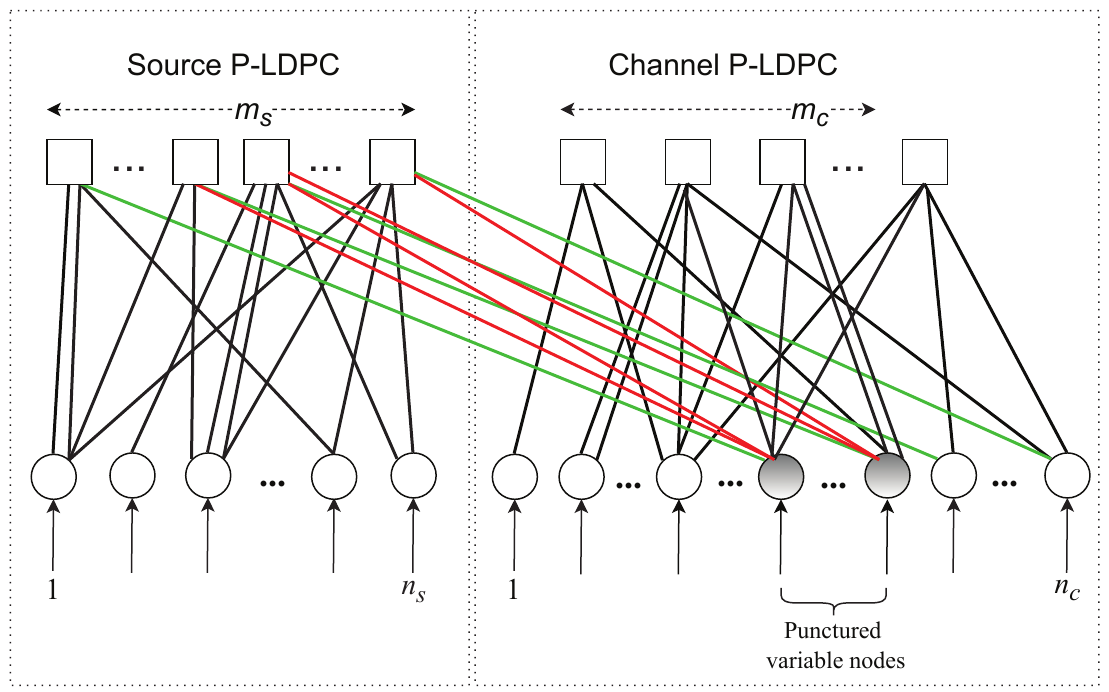}}
\caption{Representation of the traditional (without the red connections) and new class of 
(with the red connections)  DP-LDPC JSCCs. 
%Compared with the original DP-LDPC JSCC shown in Fig.~\ref{JSCC_DPLDPC}, 
%the new class of DP-LDPC JSCC allows extra connections between
%the CN set in the source P-LDPC code and the VN set in the channel P-LDPC, subject to a certain constraint. 
}
\label{JSCC_DPLDPC_modified}
\end{figure}

\section{New Class of DP-LDPC JSCC} \label{sect:DP-LDPC JSCC}
\subsection{Code Structure}
Fig.~\ref{JSCC_DPLDPC_modified} illustrates the structure of the new class of DP-LDPC JSCC (with the red connections). 
Compared with the traditional DP-LDPC JSCC, 
the new class of DP-LDPC JSCC allows
extra connections (denoted by red lines) between
the 
%{\color{black} check node (CN)} 
CN set in the source P-LDPC code and the
% {\color{black}variable node (VN)} 
VN set in the channel P-LDPC. In other words, the 
CNs in the source P-LDPC code and VNs in the channel P-LDPC code
are not linked in a one-to-one manner, but in a 
one-to-multiple manner.
However, we set a constraint on the new connections such that the new SCCV
linking protomatrix consists
of a zero matrix and a lower/upper triangular base matrix
with $1$'s on the diagonal. 
%In the lower/upper triangular base matrix, the elements on the diagonal equal $1$ and the other non-zero elements can take any non-negative integral values. 
In other words, the new class of DP-LDPC JSCC
%they can connect each VN in the channel protograph, which has been connected to a check node (CN) in the source protograph by a green line, to more source CNs and these CNs should be to the right of the CN already connected by the green line. This means more messages can be passed from the source CNs to the channel VNs. 
%This structure can also 
can be denoted by %a joint protomatrix, which is 
$\textbf{B}_{\mathrm J}=
\begin{pmatrix} 
\makebox[-0.6cm]{} \textbf{B}_{\mathrm s} &  \textbf{B}'_{\mathrm {sccv}} \\
\;\; \textbf{0}_{m_c\times n_s} & \makebox[-0.5cm]{}  \textbf{B}_{\mathrm c}
\end{pmatrix}$,
%\begin{equation}
%\textbf{B}_{\mathrm J}=
%\begin{pmatrix} 
%\makebox[-0.6cm]{} \textbf{B}_{\mathrm s} &  \textbf{B}'_{\mathrm {sccv}} \\
%\;\; \textbf{0}_{m_c\times n_s} & \makebox[-0.5cm]{}  \textbf{B}_{\mathrm c}
%\end{pmatrix}
%\label{eq:B_J}
%\end{equation}
where $\textbf{B}'_{\mathrm {sccv}}$ is the new SCCV linking protomatrix.
Furthermore, the structure of  $\textbf{B}'_{\mathrm {sccv}}$ can be written as 
\begin{equation}\label{sccv_new}
\textbf{B}'_{\mathrm {sccv}}=\begin{pmatrix} \textbf{0}_{m_s\times m_c} & \textbf{T}_{m_s}\end{pmatrix},
\end{equation}
where $\textbf{T}_{m_s}$ is an upper or a lower triangular matrix with size $m_s\times m_s$
and $1$'s on the diagonal.
%We illustrate the new class of  DP-LDPC JSCC 
%with $\textbf{T}_{m_s}$ being a lower triangular matrix. 
For example, if $\textbf{T}_{m_s}$ is a lower triangular matrix, it can be written as
%$\textbf{T}_{m_s}=\left( {\begin{array}{*{20}{c}}
%		1 & 0 & 0 & \cdots & 0 \\
%		{{t_{2,1}}} & 1 & 0 & \cdots & 0 \\
%		{{t_{3,1}}} & {{t_{3,2}}} & 1 & \vdots & \vdots \\
%		\vdots & \vdots & \cdots & \ddots & 0 \\
%		{{t_{{m_s},1}}} & {{t_{{m_s},2}}} & \cdots & {{t_{{m_s},{m_s} - 1}}} & 1 \\
%\end{array}} \right)$ 
$\textbf{T}_{m_s}=\left( {\begin{array}{*{20}{c}}
1 & 0 & 0 & \cdots & 0 \\
{{t_{2,1}}} & 1 & 0 & \cdots & 0 \\
%{{t_{3,1}}} & {{t_{3,2}}} & 1 & \vdots & \vdots \\
\vdots & \vdots & \cdots & \ddots & 0 \\
{{t_{{m_s},1}}} & {{t_{{m_s},2}}} & \cdots & {{t_{{m_s},{m_s} - 1}}} & 1 \\
\end{array}} \right)$,
%\begin{equation}
%\textbf{T}_{m_s}=\left( {\begin{array}{*{20}{c}}
%1 & 0 & 0 & \cdots & 0 \\
%{{t_{2,1}}} & 1 & 0 & \cdots & 0 \\
%%{{t_{3,1}}} & {{t_{3,2}}} & 1 & \vdots & \vdots \\
%\vdots & \vdots & \cdots & \ddots & 0 \\
%{{t_{{m_s},1}}} & {{t_{{m_s},2}}} & \cdots & {{t_{{m_s},{m_s} - 1}}} & 1 \\
%\end{array}} \right)
%\end{equation}
where $t_{i,j}$ ($i\in\{2,\ldots,m_s\}; {\color{black}j\in\{1,\ldots,i-1\}}$) are non-negative integers.
We first lift the protomatrix $\textbf{B}_{\mathrm J}$ 
with a small lifting factor $z_1$ using the progressive-edge-growth (PEG) algorithm \cite{PEG}
to form
$\textbf{B}_{\mathrm J}^{z_1}=
\left(\begin{array} {c|c}
\makebox[-0.6cm]{} \textbf{B}_{\mathrm s}^{z_1} &  \textbf{0}_{m_sz_1\times m_cz_1} \  \textbf{T}_{m_s}^{z_1}\\
\hline
\;\; \textbf{0}_{m_cz_1\times n_sz_1} & \makebox[-0.5cm]{}  \textbf{B}_{\mathrm c}^{z_1}
\end{array}\right)$,
%\begin{equation}
%\textbf{B}_{\mathrm J}^{z_1}=
%\left(\begin{array} {c|c}
%\makebox[-0.6cm]{} \textbf{B}_{\mathrm s}^{z_1} &  \textbf{0}_{m_sz_1\times m_cz_1} \  \textbf{T}_{m_s}^{z_1}\\
%\hline
%\;\; \textbf{0}_{m_cz_1\times n_sz_1} & \makebox[-0.5cm]{}  \textbf{B}_{\mathrm c}^{z_1}
%\end{array}\right) 
%\end{equation}
where
%$\textbf{B}_{\mathrm s}^{z_1}$,
%$\textbf{B}_{\mathrm c}^{z_1}$
%and $\textbf{T}_{m_s}^{z_1}$ are matrices of appropriate sizes. 
$\textbf{B}_{\mathrm s}^{z_1}$ with size $m_sz_1\times n_sz_1$ is obtained by lifting $\textbf{B}_{\mathrm s}$;
$\textbf{B}_{\mathrm c}^{z_1}$ with size $m_cz_1\times n_cz_1$ is obtained by lifting $\textbf{B}_{\mathrm c}$; and $\textbf{T}_{m_s}^{z_1}$ with size  $m_sz_1\times m_sz_1$ is obtained by lifting $\textbf{T}_{m_s}$.
%\begin{itemize}
%\item $\textbf{B}_{\mathrm s}^{z_1}$ with size $m_sz_1\times n_sz_1$ is obtained by lifting $\textbf{B}_{\mathrm s}$;
%\item $\textbf{B}_{\mathrm c}^{z_1}$ with size $m_cz_1\times n_cz_1$ is obtained by lifting $\textbf{B}_{\mathrm c}$ ;
%\item  $\textbf{T}_{m_s}^{z_1}$ with size  $m_sz_1\times m_sz_1$ is obtained by lifting $\textbf{T}_{m_s}$.
%\end{itemize}  
%
The objective of the lifting is to eliminate all entries with values larger than $1$,  
thereby obtaining a matrix with only $0$'s and $1$'s.

Then, we lift $\textbf{B}_{\mathrm J}^{z_1}$ again with a lifting factor of $z_2$, forming a large quasi-cyclic low-density parity-check (QC-LDPC) matrix of appropriate size. 
We denote the QC-LDPC matrix by 
%\small
\begin{equation} \label{QC_H}
\textbf{H}_{\mathrm J}=
\left(\begin{array} {c|c}
\makebox[-0.6cm]{} \textbf{H}_{\mathrm s}^{z_1} &  \bm{-1}_{m_sz_1\times m_cz_1} \  \textbf{H}_{{\mathrm T}_{m_s}}^{z_1}\\
\hline
\;\; \bm{-1}_{m_cz_1\times n_sz_1} & \makebox[-0.5cm]{}  \textbf{H}_{\mathrm c}^{z_1}
\end{array}\right)  .
\end{equation} \normalsize
Each entry $h$ in $\textbf{H}_{\mathrm J}$ represents a 
circulant permutation matrix (CPM)  with size $z_2\times z_2$ obtained by cyclically right-shifting the identity matrix $\textbf{I}_{z_2}$ by $h$ columns. 
Moreover, ``$-1$'' denotes a zero matrix with size $z_2\times z_2$. 
Note also that during the lifting process, we try to maximize the girth (shortest cycle) of the resultant QC-LDPC matrix. % \cite{lau2012qc}. 

\subsection{Source Encoding}
We consider a binary independent and identically distributed (i.i.d.)  source that follows  
a Bernoulli distribution, with the probability of ``1" given by $p_1$, the probability of ``0" given by $1-p_1$. 
Given a sample source sequence $\bm{s}$ of length $1\times N_s=1\times n_s z_1 z_2$,
we divide it into $n_s z_1$ sub-sequences 
$\bm{s}_i$ ({\color{black}$i\in\{1,2,\ldots,n_s z_1\}$}) each of length $z_2$. Thus
we can re-write  $\bm{s}$ as 
$\bm{s} = (\bm{s}_1 \;  \bm{s}_2 \; \cdots \; \bm{s}_{n_s z_1} \;)$.
%\begin{equation}
%\bm{s} = (\bm{s}_1 \;  \bm{s}_2 \; \cdots \; \bm{s}_{n_s z_1} \;).
%\end{equation}
%Denoting the probability of ``1" in the source sequence by $p_1$, the probability of ``0" in the source equals $1-p_1$. Therefore, the source entropy is given by
%\begin{equation}\label{entropy}
%H(p_1) = -p_1\log_2 p_1-(1-p_1)\log_2 (1-p_1)
%\end{equation} 
%where $p_1\neq 0.5$. 
Then, we generate the compressed source sequence $\bm{u}$
of length $1\times M_s$$=1\times m_sz_1z_2$ using 
{\color{black}$\textbf{H}_{\mathrm s}^{z_1}$ and $\textbf{H}_{{\mathrm T}_{m_s}}^{z_1}$} in \eqref{QC_H}.
We divide $\bm{u}$ into ${m_s}{z_1}$ groups each of length $z_2$
and represent $\bm{u}$ as
$\bm{u} = (\bm{u}_1 \;  \bm{u}_2 \; \cdots \; \bm{u}_{m_s z_1})$.
%\begin{equation}
%\bm{u} = (\bm{u}_1 \;  \bm{u}_2 \; \cdots \; \bm{u}_{m_s z_1}).
%\end{equation}
Considering the first block row of $\textbf{H}_{\mathrm J}$ shown in \eqref{QC_H},
we have 
$\sum_{i=1}^{n_s z_1}  \bm{s}_{i} (h_s^{(1,i)})^{T} +  \bm{u}_1 = \textbf{0} %\cr
\Rightarrow  
\bm{u}_1  =  \sum_{i=1}^{n_s z_1}  \bm{s}_{i} (h_s^{(1,i)})^{T}${\color{black}, where} $h_s^{(j,i)}$ ({\color{black}$j\in\{1,2,\ldots,m_sz_1\}$; $i\in\{1,2,\ldots,n_sz_1\}$}) denotes the $(j,i)$-th CPM in $\textbf{H}_{\mathrm s}^{z_1}$.
%For the second block row, we have 
%\begin{eqnarray}
%&& \sum_{i=1}^{n_s z_1}  \bm{s}_{i} (h_s^{(2,i)})^{T} +  \bm{u}_1 (h_t^{(2,1)})^{T}  +  \bm{u}_2 = \textbf{0} \cr
%\Rightarrow &&  \bm{u}_2  =  \sum_{i=1}^{n_s z_1}  \bm{s}_{i} (h_s^{(2,i)})^{T} +  \bm{u}_1 (h_t^{(2,1)})^{T}
%\end{eqnarray}
%{\color{black}where $h_t^{(j,k)}$ denotes the $(j,k)$-th CPM in $\textbf{H}_{{\mathrm T}_{m_s}}^{z_1}$.}
%{\color{black}where $h_t^{(2,1)}$ denotes the $(2,1)$-th CPM in $\textbf{H}_{{\mathrm T}_{m_s}}^{z_1}$.}
Then, 
%In general, 
%$\bm{u}_{j}$ ($j=2,3,...,m_sz_1$) can be computed sequentially using 
%\begin{eqnarray}
%&&  \sum_{i=1}^{n_s z_1}  \bm{s}_{i} (h_s^{(j,i)})^{T} 
%+ \sum_{k=1}^{j-1} \bm{u}_{k}(h_t^{(j,k)})^{T} + \bm{u}_{j}= \textbf{0} \cr
%\Rightarrow &&   \bm{u}_{j}= \sum_{i=1}^{n_s z_1}  \bm{s}_{i} (h_s^{(j,i)})^{T} 
%+ \sum_{k=1}^{j-1} \bm{u}_{k}(h_t^{(j,k)})^{T}.
%\label{eq:m_encoding}
%\end{eqnarray}
\begin{equation}
\bm{u}_{j}= \begin{array}{c} \sum_{i=1}^{n_s z_1}  \bm{s}_{i} (h_s^{(j,i)})^{T} 
+ \sum_{k=1}^{j-1} \bm{u}_{k}(h_t^{(j,k)})^{T} \end{array},
%\bm{u}_{j}= \sum_{i=1}^{n_s z_1}  \bm{s}_{i} (h_s^{(j,i)})^{T} 
%+ \sum_{k=1}^{j-1} \bm{u}_{k}(h_t^{(j,k)})^{T}.
\label{eq:m_encoding}
\end{equation}
{\color{black}where $h_t^{(j,k)}$ denotes the $(j,k)$-th CPM in $\textbf{H}_{{\mathrm T}_{m_s}}^{z_1}$.}
%Note that $\bm{s}_i$ ($i=1,2,\ldots,n_s z_1$) {\color{black}corresponds} to the 
%($(i-1)z_2+1$)-th to $i z_2$-th
%columns in $\textbf{H}_{\mathrm s}^{z_1}$ whereas
%$\bm{u}_j$ ($j=1,2,\ldots,m_s z_1$) corresponds to the ($(j-1)z_2+1$)-th to $j z_2$-th
%columns in $\textbf{T}_{m_s}^{z_1z_2}$. 
%
%\bm{s}\left(\begin{array}{ccc}
%	h_s^{(i_t,1)} & \cdots & h_s^{(i_t,n_sz_1)}  \\
%\end{array}\right)^{T} + \sum_{j_t=1}^{i_t-1} \bm{u}_{j_t}(h_t^{(i_t,j_t)})^{T}
% $\bm{u}_j$ ($i_t=2,\ldots,m_sz_1$) is 
%denoted by $\bm{u}$ with size $1\times M_s$$=1\times m_sz_1z_2$. Based on $\textbf{H}_{\mathrm J}$ shown in \eqref{mdp-jscc2}, we can divide $\bm{u}$ into $m_sz_1$ groups and each group has $z_2$ compressed source symbols. Then we set  $\bm{u}=\{\bm{u}_1,\bm{u}_2,...,\bm{u}_{m_sz_1}\}$ and use $\bm{u}_{i_t}$ ($i_t=1,2,...,m_sz_1$) to denote the $i$-th group of $\bm{u}$. We can first calculate $\bm{u}_1$ by using 
%$\bm{u}_1=\bm{s}\left(\begin{array}{ccc}
%	h_s^{(1,1)} & \cdots & h_s^{(1,n_sz_1)}  \\
%\end{array}\right)^{T}$.
%
%
%Then we calculate $\bm{u}_{i_t}$ ($i_t=2,3,...,m_sz_1$) sequentially by 
%\begin{equation}
%\bm{u}_{i_t}=\bm{s}\left(\begin{array}{ccc}
%	h_s^{(i_t,1)} & \cdots & h_s^{(i_t,n_sz_1)}  \\
%\end{array}\right)^{T} + \sum_{j_t=1}^{i_t-1} \bm{u}_{j_t}(h_t^{(i_t,j_t)})^{T}
%\end{equation}
%where $+$ here means the modulo $2$ addition. 
%Since CPMs are used, the $z_2$ values in $\bm{u}_{j}$ (${\color{black}j}=1,2,...,m_sz_1$)
%can be evaluated in parallel using simple shift and xor operations.
%A maximum degree of parallelism is therefore $z_2$ for the 
%new class of DP-LDPC JSCC. 
Afterwards, the compressed source symbols $\bm{u}$ 
is passed to the original channel coding to evaluate
the parity-check bits based on  {\color{black}$ \textbf{H}_{\mathrm c}^{z_1}$}.

{\color{black}
\subsection{Complexity and Latency Analyses}\label{sect:analysis}
\subsubsection{Source encoder}
In the traditional DP-LDPC JSCC system, $\bm{u}_{j}$ ({\color{black}$j\in\{1,2,\ldots,m_sz_1\}$}) can be computed in full parallel using 
\begin{equation}
	\label{eq:trad_encoding}
	\bm{u}_{j} = 
	\begin{array}{cc}
		\sum_{i=1}^{n_s z_1}  \bm{s}_{i} (h_s^{(j,i)})^{T}, & {\color{black}j\in\{1,2,\ldots,m_sz_1\}} \\	
	\end{array},
\end{equation}
%		In a practical environment, however, $\bm{u}$ would not be derived in full parallel 
%		because 
but it requires a lot of hardware resources. 
The most common implementation %of the hardware
%source encoder 
is to compute $\bm{u}_{j}$ sequentially, i.e., $\bm{u}_{1}$, $\bm{u}_{2}$, ...,
$\bm{u}_{m_sz_1}$. 
Both \eqref{eq:m_encoding} and \eqref{eq:trad_encoding} above can be completed with simple shift registers and {\color{black}XOR} gates.
Thus, the {source encoding complexity} of both systems {\color{black}is} considered as {low} 
though the proposed system could be relatively more complex.
Assuming a balanced binary tree structure is used to 
compute \eqref{eq:m_encoding} or \eqref{eq:trad_encoding},
it can be shown that the number of {\color{black}XOR} gates required in the source encoding process is proportional to the largest row weight of $(\textbf{B}_{\mathrm s} \  \textbf{B}_{\mathrm {sccv}})$  minus $2$ for the traditional DP-LDPC code, or the largest row weight of $(\textbf{B}_{\mathrm s} \  \textbf{B}'_{\mathrm {sccv}}) $ minus $2$ for the proposed DP-LDPC code.
Moreover, we denote $w_s^{j}$ and $w_t^{j}$
as the $j$-th row {\color{black}weights} of $\textbf{B}_{\mathrm s}^{z_1}$ and $\textbf{T}_{{m_s}}^{z_1}$, respectively.
Then the latency for encoding each $\bm{u}_{j}$ can be shown proportional to 
$\lceil\log_2 w_s^{j}\rceil$ and
$\lceil\log_2 (w_s^{j}+w_t^{j}-1)\rceil$
for traditional DP-LDPC code and the proposed DP-LDPC code, respectively,
Therefore, the percentage increase in source encoding latency is given by
\begin{equation}
	\Delta_{\rm latency,~source}= \frac{\sum_j \lceil\log_2 (w_s^{j}+w_t^{j}-1)\rceil  -\sum_j \lceil\log_2 w_s^{j}\rceil}{\sum_j \lceil\log_2 w_s^{j}\rceil }.
	\label{eq:Delta_latency}
\end{equation}

\subsubsection{Decoder}
As in 
\cite{Chen2019,liu2020joint}, we assume that a standard belief propagation (BP) algorithm is used to decode
the DP-LDPC code as one single code (i.e., $\textbf{B}_{\mathrm J_0}$ or $\textbf{B}_{\mathrm J}$ after lifting is considered), and 
to update the check-to-variable (C2V) messages and variable-to-check (V2C) messages iteratively.
Since the computation of C2V messages is much more complex than that of V2C messages,
we approximate the complexity of the decoder by the complexity of the check-node processors (CNPs).
We further assume using the symmetric binary tree structure in \cite{lu20163} to compute C2V messages
during the hardware implementation, and using 
the look-up table (LUT) method to implement the ``$\tanh$'' function.
Compared with the original DP-LDPC JSCC scheme, the percentage increase in the decoding latency of the new DP-LDPC JSCC scheme can be shown equal to
\begin{equation}
	\Delta_{\rm latency,~dec} = %\frac{\sum_j 2(\lceil\log_2 w_{J}^{j}\rceil -1)  - \sum_j 2(\lceil\log_2 w_{J_0}^{j}\rceil -1)}{\sum_j 2(\lceil\log_2 w_{J_0}^{j}\rceil -1)} 
	%	= 
	\frac{\sum_j (\lceil\log_2 w_{J}^{j}\rceil - \lceil\log_2 w_{J_0}^{j}\rceil )}{\sum_j \lceil\log_2 w_{J_0}^{j}\rceil -1},
	\label{eq:latency}
\end{equation}
where $w_{J}^{j}$ and $w_{J_0}^{j}$ denote the $j$-th row weights of $\textbf{B}_{\mathrm J}$ and $\textbf{B}_{\mathrm J_0}$, respectively.
Similar to the source encoder complexity, 
the complexity of the CNP depends on the 
CN with the highest degree. When the highest row weight is $x$, $3(x-2)$ LUTs {\color{black}are} needed to implement  the symmetric binary tree structure.}

\section{Results and discussions}\label{sect:results}
%\subsection{Example \#1}
\noindent \underline{Example \#1:} We consider the DP-LDPC code in
\cite[Eq.(16)]{liu2020joint} which is designed at $p_1=0.04$. We denote it by
${\textbf{B}}_{\mathrm J}^{0.04} = {\textbf{B}}^{0.04} (x_1=x_2=0)${\color{black}, where} ${\textbf{B}}^{0.04}$ is given as %\eqref{BJopti_1}.
%\small
\begin{equation}\label{BJopti_1}
	\begin{array}{l}
		{\textbf{B}}^{0.04} %{\textbf{B}}_{\mathrm J}^{0.04}
		= \left( {\begin{array}{cccc|ccccc}
				2 & 2 & 1 & 1 & 0 & 0 & 0 & 1 & x_2\\
				1 & 1 & 2 & 1 & 0 & 0 & 0 & x_1 & 1\\
				\hline
				0 & 0 & 0 & 0 & 1 & 0 & 1 & 2 & 2\\
				0 & 0 & 0 & 0 & 0 & 1 & 1 & 1 & 1\\
				0 & 0 & 0 & 0 & 0 & 1 & 1 & 0 & 2\\
		\end{array}} \right) \\ 
	\end{array}.
\end{equation}
\normalsize
For ${\textbf{B}}_{\mathrm J}^{0.04}$, $m_s=2,n_s=4,m_c=3,n_c=5$ and the last VN  
in the channel protograph is punctured. 
The overall symbol code rate of this DP-LDPC JSCC is evaluated by 
$R = R_s R_c = 1${\color{black}, where} $R_s=n_s/m_s$ is the source compression rate, $R_c=m_s/(n_c-n_p)$ denotes the channel code rate, and $n_p$ denotes the number of punctured VNs in the channel protograph. {\color{black} By using the JPEXIT algorithm \cite{chen2016performance,Chen2019,liu2020joint}, the channel threshold of ${\textbf{B}}_{\mathrm J}^{0.04}$ is found to be $-5.127$ dB.}

\begin{table*}[!t]
\centering
\caption{The channel thresholds $(E_s/N_0)_{\rm T}~(\rm dB)$ of $\textbf{B}^{0.04}$ in \eqref{BJopti_1} for different $x_1$ and $x_2$ values. The Shannon limit is $-7.0~{\rm dB}$.}
\begin{tabular}{|c||c|c|c||c||c|c|c|}
	\hline
	$x_1$ & $0$  & $0$ & $0$   & $0$ &  $1$ & $2$ & $3$ \\
	\hline	
	$x_2$ & $1$  & $2$ & $3$   & $0$ &  $0$ & $0$ & $0$ \\
	\hline
	$(E_s/N_0)_{\rm T}$ (dB) & $\bm{-5.267}$ & $-5.204$ & $-5.049$ & $-5.127$ & $-4.819$ & $-4.526$ & $-4.273$ \\
	%\hline		
	%\makecell{Shannon limit (dB)}& \multicolumn{8}{c|}{$-12.02$} \\
	\hline
\end{tabular}
\label{Bopt_0p04_diff}
\end{table*}

%It is designed in \cite{liu2020joint} when $R=1$ and $p_1=0.04$. We display it in \eqref{BJopti_1} and use $\textbf{B}_{\mathrm J}^{0.04}$ to denote it in this paper. The fifth VN with the highest degree in the channel code of $\textbf{B}_{\mathrm J}^{0.04}$ is punctured. 
In this paper, we assume the maximum entry value of a code is $3$ to limit the searching space. To construct a new DP-LDPC code based on ${\textbf{B}}_{\mathrm J}^{0.04}$, we consider
%	 two cases: (i) $x_1$ is a positive integer and $x_2=0$, i.e., 
%	$\begin{pmatrix} 1 & x_2 \\ x_1 & 1 \end{pmatrix}$ is a lower triangular matrix; 
%	and (ii) $x_1=0$ and $x_2$ is a positive integer, i.e., $\begin{pmatrix} 1 & x_2 \\ x_1 & 1 \end{pmatrix}$
%	is an upper triangular matrix. 
%	We consider 
all possible lower and upper triangular structures of $\begin{pmatrix} 1 & x_2 \\ x_1 & 1 \end{pmatrix}$ and evaluate the channel thresholds of the corresponding codes.
We can see from  Table \uppercase\expandafter{\romannumeral1} 
that when $x_1=0$ and $x_2=1$, $\textbf{B}^{0.04}$ has the lowest channel threshold. 
We denote this optimized code by ${\textbf{B}}_{\mathrm J\_{\rm opt1}}^{0.04}= {\textbf{B}}^{0.04} (x_1=0,x_2=1)$.
% Its corresponding code is denoted by	
%\begin{equation}\label{BJopti_1_opt1}
%\begin{array}{l}
%	{\textbf{B}}_{\mathrm J\_{\rm opt1}}^{0.04}
%	= \left( {\begin{array}{cccc|ccccc}
	%			2 & 2 & 1 & 1 & 0 & 0 & 0 & 1 & {\color{black}1}\\
	%			1 & 1 & 2 & 1 & 0 & 0 & 0 & 0 & 1\\
	%			\hline
	%			0 & 0 & 0 & 0 & 1 & 0 & 1 & 2 & 2\\
	%			0 & 0 & 0 & 0 & 0 & 1 & 1 & 1 & 1\\
	%			0 & 0 & 0 & 0 & 0 & 1 & 1 & 0 & 2\\
	%	\end{array}} \right) \\ 
%\end{array}.	
%\end{equation}	
The channel threshold of ${\textbf{B}}_{\mathrm J\_{\rm {opt1}}}^{0.04}$ is found to be 
$-5.267$ dB, which is $0.140$ dB lower than the threshold of $\textbf{B}_{\mathrm J}^{0.04}$. % {\color{black}where $x_1=x_2=0$}. 
We also run simulations and record the source symbol error rate (SSER) of the code. 
In  {\color{black}all} the simulation results, we set the maximum number of decoding iterations to $I_{\max}=200$. 
The simulation will be terminated if (i) more than $2\times10^{5}$  frames have been simulated or (ii) more than $100$ error frames have been found and no less than $5000$ frames have been simulated. 
Fig.~\ref{letter_db_res3} shows the simulated results for ${\textbf{B}}_{\mathrm J}^{0.04}$ and ${\textbf{B}}_{\mathrm J\_{\rm {opt1}}}^{0.04}$. We can see that at {\color{black}an} SSER of $10^{-6}$, ${\textbf{B}}_{\mathrm J\_{\rm {opt1}}}^{0.04}$ outperforms ${\textbf{B}}_{\mathrm J}^{0.04}$ by about $0.25$ dB.
%
%{\color{black}
%Since the largest row weights of $(\textbf{B}_{\mathrm s} \  \textbf{B}_{\mathrm {sccv}})$ in ${\textbf{B}}_{\mathrm J}^{0.04}$ 
%and $(\textbf{B}_{\mathrm s} \  \textbf{B}'_{\mathrm {sccv}})$ in  ${\textbf{B}}_{\mathrm J\_{\rm opt1}}^{0.04}$ are
%$7$ and $8$, respectively, the percentage increase in 
%the number of xor gates used in source encoding equals $\Delta_{\rm xor}=[(8-2)-(7-2)]/(7-2)
%= 20\%$. 
%Even though the percentage increase in the source encoding complexity
%may seem large, the increase in the overall 
%encoder complexity 
%(when considering both source encoding and channel encoding) 
%is much lower. 
%The percentage increase in 
%the number of LUTs used in C2V updates (decoder complexity) equals $(3(8-2)-3(7-2))/(3(7-2))
%= 20\%$. 
%However, there is no change in the source encoding latency 
%or DP-LDPC decoding latency according to the analyses in Sect.~\ref{sect:analysis}
%% because 
%%$(\lceil\log_2 (8-1)\rceil-\lceil\log_2 (7-1)\rceil) = 0$.  
%}

\begin{figure}[htp]
\centering
{\includegraphics[width=2.4in]{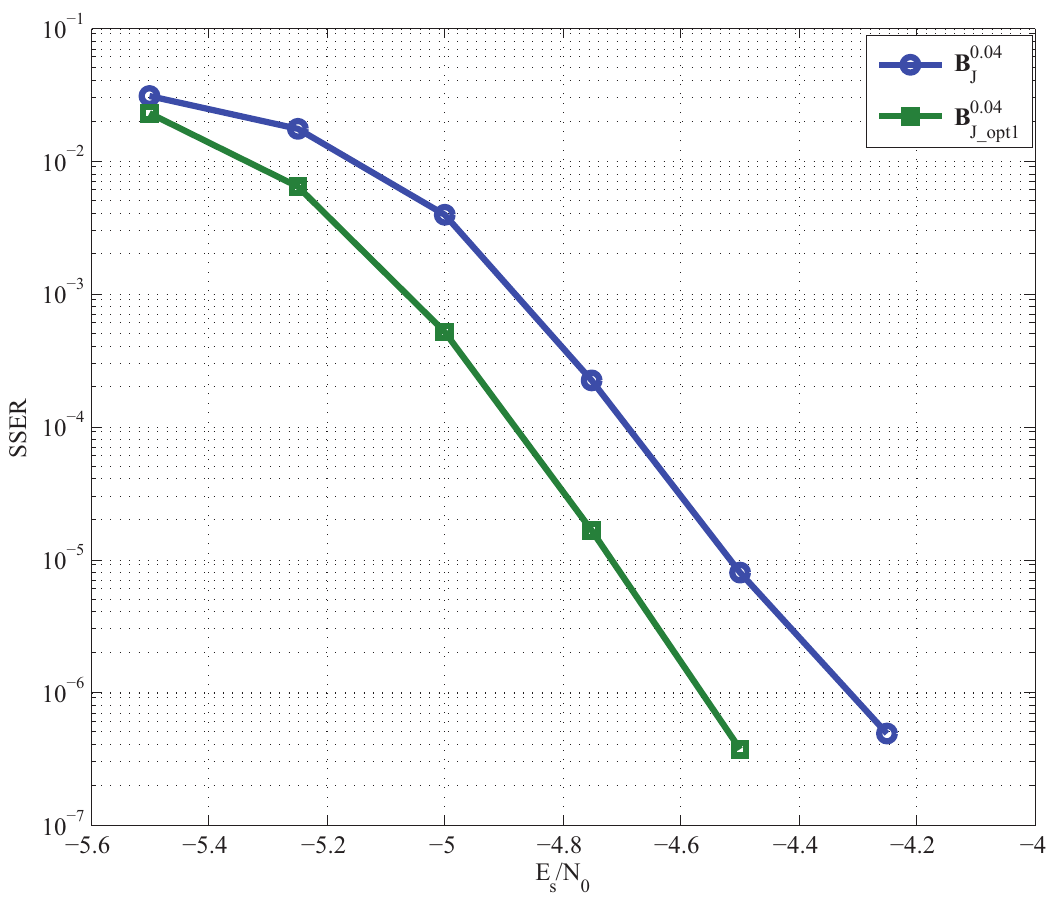}}
\caption{SSER performance comparison for ${\textbf{B}}_{\mathrm J}^{0.04}$ and ${\textbf{B}}_{\mathrm {J\_{opt1}}}^{0.04}$ when $R=1$, $p_1=0.04$, and $N_s=z_1z_2n_s=4\times800\times4=12800$. {\color{black}The Shannon limit is $-7.00$ dB.}	
}
\label{letter_db_res3}
\end{figure}
%We take the DP-LDPC JSCCs shown in \cite{Chen2019} and \cite{liu2020joint} as examples to show the feasibility and effectiveness of our proposed new class of DP-LDPC JSCCs. 

%\subsection{Example \#2} 	
\noindent \underline{Example \#2:} \footnote{\label{note1}{\color{black}We swap the fourth and fifth columns of the SCCV linking matrices of $\textbf{B}_{\mathrm J3}^{0.01}$ and $\textbf{B}_{\mathrm J4}^{0.01}$ in \cite[Table \uppercase\expandafter{\romannumeral1}]{Chen2019} to obtain an identity matrix. Correspondingly, the fourth and fifth columns of the channel base matrices of these two codes are also swapped. 
%So, in this paper, the channel base matrices and SCCV linking matrices of $\textbf{B}_{\mathrm J3}^{0.01}$ and $\textbf{B}_{\mathrm J4}^{0.01}$ are respectively shown in \eqref{BJ3}, \eqref{BJ4} and \eqref{sccv}.
}} We consider two DP-LDPCs $\textbf{B}_{\mathrm J3}^{0.01}$ and $\textbf{B}_{\mathrm J4}^{0.01}$ in \cite[Table \uppercase\expandafter{\romannumeral1}]{Chen2019} 
which are designed for $p_1=0.01$ and $R=2$. For $\textbf{B}_{\mathrm J3}^{0.01}$,
the source base matrix ${\textbf{B}}_{\mathrm {J3-s}}^{0.01}$ and channel base matrix ${\textbf{B}}_{\mathrm {J3-c}}^{0.01}$ 
are given by 

\begin{equation}\label{BJ3}
%\small
\left\{\begin{aligned}
&\begin{array}{l}
	{\textbf{B}}_{\mathrm {J3-s}}^{0.01}
	= \left( {\begin{array}{*{20}{c}}
			1 & 1 & 2 & 1 & 3 & 1 & 3 & 1\\
			1 & 2 & 1 & 2 & 1 & 2 & 1 & 2\\
	\end{array}} \right) \\ 
\end{array}
\\
&\begin{array}{l} {\textbf{B}}_{\mathrm {J3-c}}^{0.01}
	= \left( {\begin{array}{*{20}{c}}
			1 & 0 & 0 & {\color{black}3} & {\color{black}0} \\
			0 & 1 & 1 & {\color{black}1} & {\color{black}2} \\
			0 & 1 & 1 & {\color{black}2} & {\color{black}1} \\
	\end{array}} \right)%^{\color{black}\footref{note1}}
	\\ 
\end{array} 
\end{aligned}
\right. .
\end{equation}
For $\textbf{B}_{\mathrm J4}^{0.01}$,
the source base matrix ${\textbf{B}}_{\mathrm {J4-s}}^{0.01}$ and channel base matrix ${\textbf{B}}_{\mathrm {J4-c}}^{0.01}$
are given by 
\begin{equation}\label{BJ4}
%\small
\left\{\begin{aligned}
&\begin{array}{l}
	{\textbf{B}}_{\mathrm J4-s}^{0.01}
	= \left( {\begin{array}{*{20}{c}}
			2 & 1 & 2 & 1 & 3 & 1 & 3 & 1\\
			1 & 2 & 1 & 2 & 1 & 3 & 1 & 3\\
	\end{array}} \right) \\ 
\end{array}
\\
&\begin{array}{l}
	{\textbf{B}}_{\mathrm J4-c}^{0.01}
	= \left( {\begin{array}{*{20}{c}}
			1 & 0 & 0 & {\color{black}3} & {\color{black}0} \\
			0 & 1 & 1 & {\color{black}1} & {\color{black}1} \\
			0 & 1 & 1 & {\color{black}2} & {\color{black}1} \\
	\end{array}} \right) %^{\color{black}\footref{note1}} 
	\\ 
\end{array} \\
\end{aligned}
\right. .
\end{equation}
For both $\textbf{B}_{\mathrm J3}^{0.01}$ and $\textbf{B}_{\mathrm J4}^{0.01}$, $m_s=2,n_s=8,m_c=3,n_c=5$, and $n_p=1$. The punctured VN corresponds to the {\color{black}fourth} VN in the channel protograph. 
Their SCCV linking base matrices are both %given by
${\textbf{B}}_{\mathrm {J3-sccv}}^{0.01} = {\textbf{B}}_{\mathrm {J4-sccv}}^{0.01} = {\textbf{B}}_{\mathrm {sccv}}^{0.01} (x_1=x_2=0)${\color{black}, where}
\begin{equation}\label{sccv}
%\small
{\textbf{B}}_{\mathrm {sccv}}^{0.01} = 	\begin{array}{l}
	\left( {\begin{array}{*{20}{c}}
			0 & 0 & 0 & 1 & x_2 \\
			0 & 0 & 0 & x_1 & 1 \\
	\end{array}} \right) %^{\color{black}\footref{note1}}
	\\ 
\end{array}.
\end{equation}
%where $x_1=x_2=0$.
%
% $\textbf{B}_{\mathrm J3-s}^{0.01}$ and $\textbf{B}_{\mathrm J3-c}^{0.01}$ shown in \eqref{BJ3} are the source and channel protomatrices of $\textbf{B}_{\mathrm J3}^{0.01}$, respectively. $\textbf{B}_{\mathrm J4-s}^{0.01}$ and $\textbf{B}_{\mathrm J4-c}^{0.01}$ shown in \eqref{BJ4} are the source and channel protomatrices of $\textbf{B}_{\mathrm J4}^{0.01}$, respectively. $\textbf{B}_{\mathrm J3-c}^{0.01}$ and $\textbf{B}_{\mathrm J4-c}^{0.01}$ both have one punctured VN and it is the fifth VN. The source-check-channel-variable (SCCV) linking protomatrices of $\textbf{B}_{\mathrm J3}^{0.01}$ and $\textbf{B}_{\mathrm J4}^{0.01}$ are the same and they are shown in \eqref{B_sccv}. 
%
%
%
%\begin{equation} \label{B_sccv} 
%\begin{pmatrix} \textbf{0}_{m_s\times m_c} & \textbf{I}_{m_s}\end{pmatrix}=\begin{array}{l}
%\left( {\begin{array}{*{20}{c}}
%0 & 0 & 0 & 0 & 1 \\
%0 & 0 & 0 & 1 & 0 \\
%\end{array}} \right) \\ 
%\end{array}.
%\end{equation} 
%		As in Example \#1, we {\color{black}set either $x_1$ or $x_2$ to a positive integer ($\leq 3$) to obtain a lower or upper triangular matrix. By using the JPEXIT algorithm and searching all the possibilities, 
Using the same method as in Example \#1, we find three cases, i.e., 
	${\textbf{B}}_{\mathrm {sccv}}^{0.01}(x_1=1,x_2=0)$, ${\textbf{B}}_{\mathrm {sccv}}^{0.01}(x_1=2,x_2=0)$, and 
	${\textbf{B}}_{\mathrm {sccv}}^{0.01}(x_1=3,x_2=0)$, for which the constructed codes have lower channel thresholds than the original ones. We further denote the corresponding
	%	
	%	They are a). $x_1=1$ and $x_2=0$; b). $x_1=2$ and $x_2=0$;  c). $x_1=3$ and $x_2=0$, respectively. The corresponding new SCCV linking protomatrices are respectively denoted by} 
%
%	\begin{equation} \label{B_sccv_opt1} 
	%		\textbf{B}_{sccv1}^{'}=
	%		\begin{pmatrix} \textbf{0}_{m_s\times m_c} & \textbf{T}_{m_s}\end{pmatrix}=\begin{array}{l}
		%			\left( {\begin{array}{*{20}{c}}
				%					0 & 0 & 0 & \color{black}1 & \color{black}0 \\
				%					0 & 0 & 0 & \color{black}1 & \color{black}1 \\
				%			\end{array}} \right) \\ 
		%		\end{array}
	%	\end{equation} 
%	
%	\begin{equation} \label{B_sccv_opt2} 
	%		\textbf{B}_{sccv2}^{'}=\begin{pmatrix} \textbf{0}_{m_s\times m_c} & \textbf{T}_{m_s}\end{pmatrix}=\begin{array}{l}
		%			\left( {\begin{array}{*{20}{c}}
				%					0 & 0 & 0 & \color{black}1 & \color{black}0 \\
				%					0 & 0 & 0 & \color{black}2 & \color{black}1 \\
				%			\end{array}} \right) \\ 
		%		\end{array}
	%	\end{equation} 
%	
%	\begin{equation} \label{B_sccv_opt3} 
	%		\textbf{B}_{sccv3}^{'}=\begin{pmatrix} \textbf{0}_{m_s\times m_c} & \textbf{T}_{m_s}\end{pmatrix}=\begin{array}{l}
		%			\left( {\begin{array}{*{20}{c}}
				%					0 & 0 & 0 & \color{black}1 & \color{black}0 \\
				%					0 & 0 & 0 & \color{black}3 & \color{black}1 \\
				%			\end{array}} \right) \\ 
		%		\end{array}
	%	\end{equation} 
%We can optimize $\textbf{B}_{\mathrm J3}^{0.01}$ and $\textbf{B}_{\mathrm J4}^{0.01}$ by turning their SCCV linking matrix shown in \eqref{B_sccv} into the protomatrices shown in \eqref{B_sccv_opt1}, \eqref{B_sccv_opt2} and \eqref{B_sccv_opt3}, respectively. 
%The 
%
new DP-LDPC codes by
%	corresponding to 
%	 $\textbf{B}_{sccv1}^{'}$,
%	$\textbf{B}_{sccv2}^{'}$ and 
%	$\textbf{B}_{sccv3}^{'}$ 
%	are denoted by 
$\textbf{B}_{\mathrm {J3(J4)\_{opt1}}}^{0.01}$, $\textbf{B}_{\mathrm {J3(J4)\_{opt2}}}^{0.01}$, and $\textbf{B}_{\mathrm {J3(J4)\_{opt3}}}^{0.01}$, respectively.
%
%
%based on $\textbf{B}_{\mathrm J3}^{0.01}$ by using this method are defined as $\textbf{B}_{\mathrm J3\_{\rm opt1}}^{0.01}$, $\textbf{B}_{\mathrm J3\_{\rm opt2}}^{0.01}$ and $\textbf{B}_{\mathrm J3\_{\rm opt3}}^{0.01}$, respectively. The new DP-LDPC JSCCs obtained based on $\textbf{B}_{\mathrm J4}^{0.01}$ by using the same method are defined as $\textbf{B}_{\mathrm J4\_{\rm opt1}}^{0.01}$, $\textbf{B}_{\mathrm J4\_{\rm opt2}}^{0.01}$ and $\textbf{B}_{\mathrm J4\_{\rm opt3}}^{0.01}$, respectively. 
%	The channel thresholds of the original and new class of DP-LDPC codes are listed 
%with the original 
%DP-LDPC JSCC 
%$\textbf{B}_{\mathrm J3}^{0.01}$, $\textbf{B}_{\mathrm J4}^{0.01}$ and new DP-LDPC JSCCs generated based on them are shown 
In Table \ref{opt_tab1},
we can see that the new class of DP-LDPC codes has better thresholds compared with 
the original ones. 
%generated based on $\textbf{B}_{\mathrm J3}^{0.01}$ have lower channel thresholds than $\textbf{B}_{\mathrm J3}^{0.01}$, and among them, 
For example, the  channel threshold 
of $\textbf{B}_{\mathrm J3\_{\rm opt3}}^{0.01}$ is $0.41$ dB lower than that of $\textbf{B}_{\mathrm J3}^{0.01}$;
and the channel threshold 
of $\textbf{B}_{\mathrm J4\_{\rm opt3}}^{0.01}$ is $0.354$ dB lower than that of $\textbf{B}_{\mathrm J4}^{0.01}$.	
%All new DP-LDPC JSCCs generated based on $\textbf{B}_{\mathrm J4}^{0.01}$ also have lower channel thresholds than $\textbf{B}_{\mathrm J4}^{0.01}$, and among them, $\textbf{B}_{\mathrm J4\_{\rm opt3}}^{0.01}$ has the lowest channel threshold, which is $0.354$ dB lower than the threshold of $\textbf{B}_{\mathrm J4}^{0.01}$. 

\begin{table*}[!t]
\caption{The channel thresholds $(E_s/N_0)_{\rm T}$ of DP-LDPCs. $R=2$ and $p_1=0.01$. {\color{black}The Shannon limit is $-12.02~{\rm dB}$.}}		
\centering
\begin{tabular}{|c|c|c|c|c||c|c|c|c|}
	\hline
	~ & ${\textbf{B}}_{\mathrm J3}^{0.01}$  & ${\textbf{B}}_{\mathrm J3\_{\rm opt1}}^{0.01}$ & ${\textbf{B}}_{\mathrm J3\_{\rm opt2}}^{0.01}$   & ${\textbf{B}}_{\mathrm J3\_{\rm opt3}}^{0.01}$ &  ${\textbf{B}}_{\mathrm J4}^{0.01}$ & ${\textbf{B}}_{\mathrm J4\_{\rm opt1}}^{0.01}$ & ${\textbf{B}}_{\mathrm J4\_{\rm opt2}}^{0.01}$ & ${\textbf{B}}_{\mathrm J4\_{\rm opt3}}^{0.01}$\\
	\hline	
	$(E_s/N_0)_{\rm T}$ (dB) & $-9.324$ & $-9.555$ & $-9.680$ & $-9.734$ & $-9.390$ & $-9.616$ & $-9.722$ & $-9.744$  \\
	%\hline		
	%\makecell{Shannon limit (dB)}& \multicolumn{8}{c|}{$-12.02$} \\
	\hline
\end{tabular}
\label{opt_tab1}
\end{table*}

%\begin{table*}[!t]
%	\caption{Increased complexity of the new class of DP-LDPC JSCCs.}	
%	\centering
%	\begin{tabular}{|c|c|c|c|c||c|c|c|c|}
%		\hline
%		~ & ${\textbf{B}}_{\mathrm J3\_{\rm opt1}}^{0.01}$/${\textbf{B}}_{\mathrm J4\_{\rm opt1}}^{0.01}$ & ${\textbf{B}}_{\mathrm J3\_{\rm opt2}}^{0.01}$/${\textbf{B}}_{\mathrm J4\_{\rm opt2}}^{0.01}$   & ${\textbf{B}}_{\mathrm J3\_{\rm opt3}}^{0.01}$/${\textbf{B}}_{\mathrm J4\_{\rm opt3}}^{0.01}$ &   ${\textbf{B}}_{\mathrm J4\_{\rm opt1}}^{0.01}$ & ${\textbf{B}}_{\mathrm J4\_{\rm opt2}}^{0.01}$ & ${\textbf{B}}_{\mathrm J4\_{\rm opt3}}^{0.01}$\\
%		\hline	
%		$(E_s/N_0)_{\rm T}$ (dB) & $-9.324$ & $-9.555$ & $-9.680$ & $-9.734$ & $-9.390$ & $-9.616$ & $-9.722$ & $-9.744$  \\
%		%\hline		
%		%\makecell{Shannon limit (dB)}& \multicolumn{8}{c|}{$-12.02$} \\
%		\hline
%	\end{tabular}
%	\label{opt_tab1}
%\end{table*}

%\begin{table*}[!t]
%\caption{The channel thresholds of codes at $R=1$ and $p_1=0.04$}	
%\centering
%\begin{tabular}{|c|c|c|}
%\hline
%~ & ${\textbf{B}}_{\mathrm J}^{0.04}$  & ${\textbf{B}}_{\mathrm J\_{\rm opt1}}^{0.04}$ \\
%\hline	
%$(E_s/N_0)_{\rm T}$ (dB) & $-5.127$ & $-5.267$ \\
%\hline		
%\makecell{Shannon limit (dB)}& \multicolumn{2}{c|}{$-7.00$} \\
%\hline
%\end{tabular}
%\label{opt_tab2}
%\end{table*}

\begin{figure}[htp]
{%\color{black}
	\centering
	{\includegraphics[width=2.4in]{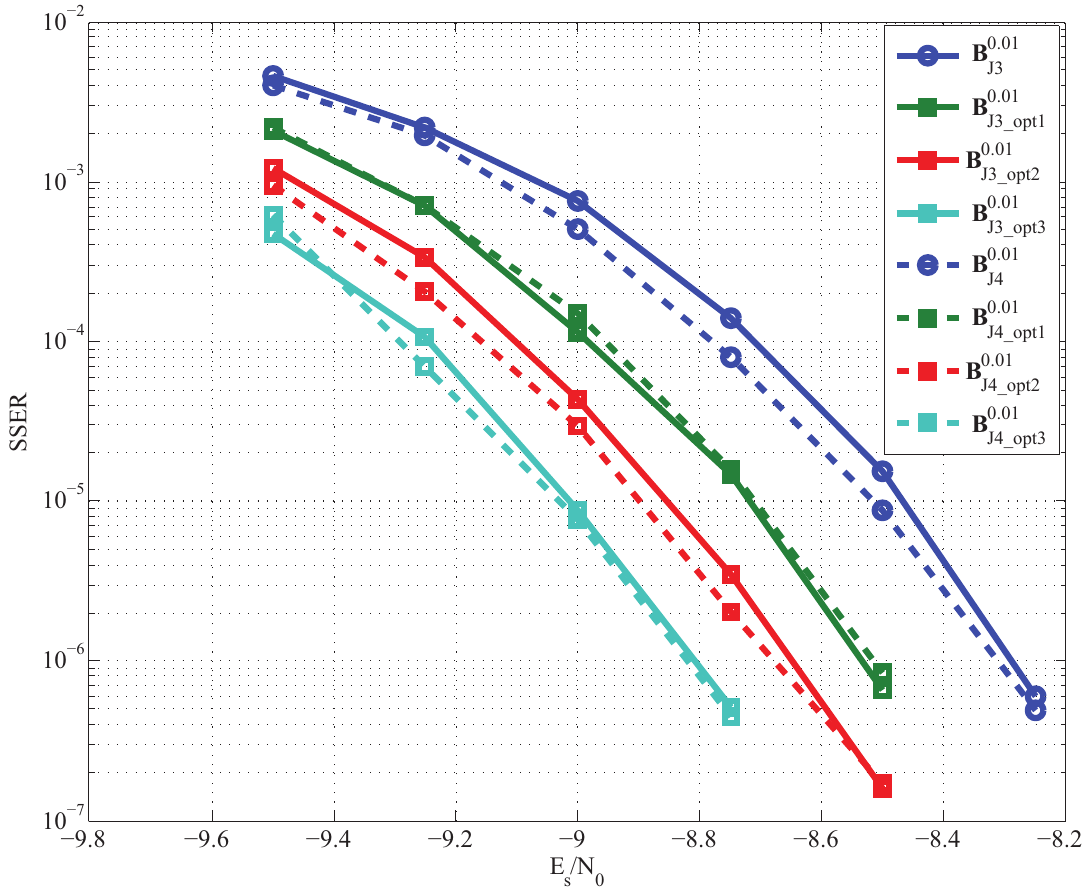}}
	\caption{SSER performance of ${\textbf{B}}_{\mathrm J3}^{0.01}$ (solid line) and ${\textbf{B}}_{\mathrm J4}^{0.01}$ (dashed line) and their corresponding new DP-LDPC codes. $R=2$, $p_1=0.01$ and $N_s=z_1z_2n_s=4\times400\times8=12800$. {\color{black}The Shannon limit is $-12.02$ dB.}} 	\label{letter_db_res1}}
\end{figure}

%	\begin{figure}[htp]
%		\centering
%		{\includegraphics[width=2.8in]{letter_db_res2.eps}}
%		\caption{SSER performance of ${\textbf{B}}_{\mathrm J4}^{0.01}$ and the new class of DP-LDPC JSCC. 
	%			$R=2$, $p_1=0.01$ and $N_s=z_1z_2n_s=4\times400\times8=12800$. 
	%		}
%		\label{letter_db_res2}
%	\end{figure}

The simulated SSERs in Fig.~\ref{letter_db_res1} show that the new class of DP-LDPC JSCC outperforms the original DP-LDPC JSCC. 
The SSER results are also consistent with the channel threshold results shown in Table \ref{opt_tab1}. 
For example, in terms of SSER, ${\textbf{B}}_{\mathrm J3\_{\rm opt3}}^{0.01}$ outperforms ${\textbf{B}}_{\mathrm J3\_{\rm opt2}}^{0.01}$, which outperforms ${\textbf{B}}_{\mathrm J3\_{\rm opt1}}^{0.01}$, which in turn 
outperforms ${\textbf{B}}_{\mathrm J3}^{0.01}$. 
When comparing the channel thresholds, 
${\textbf{B}}_{\mathrm J3\_{\rm opt3}}^{0.01} < {\textbf{B}}_{\mathrm J3\_{\rm opt2}}^{0.01} < {\color{black}{\textbf{B}}_{\mathrm J3\_{\rm opt1}}^{0.01}} < {\textbf{B}}_{\mathrm J3}^{0.01}$.
In particular, 
%Fig.~\ref{letter_db_res2} shows SSER simulation results of ${\textbf{B}}_{\mathrm J4}^{0.01}$,  ${\textbf{B}}_{\mathrm J4\_{\rm opt1}}^{0.01}$, ${\textbf{B}}_{\mathrm J4\_{\rm opt2}}^{0.01}$ and ${\textbf{B}}_{\mathrm J4\_{\rm opt3}}^{0.01}$. We can see that all new DP-LDPC JSCCs have better error performance than original DP-LDPC JSCCs. This matches the channel threshold results. 
${\textbf{B}}_{\mathrm J3\_{\rm opt3}}^{0.01}$ and ${\textbf{B}}_{\mathrm J4\_{\rm opt3}}^{0.01}$ have around $0.5$ dB coding gains over ${\textbf{B}}_{\mathrm J3}^{0.01}$ and ${\textbf{B}}_{\mathrm J4}^{0.01}$, respectively, at {\color{black}an} SSER of $10^{-6}$. 

%
%{\color{black}
%For complexity and latency, we consider the case when $\textbf{B}_{\mathrm J4}^{0.01}$ is modified to $\textbf{B}_{\mathrm J4\_{\rm opt3}}^{0.01}$. 
%The largest row weight of $(\textbf{B}_{\mathrm s} \  \textbf{B}_{\mathrm {sccv}})$ is $15$ while that of 
%$(\textbf{B}_{\mathrm s} \  \textbf{B}'_{\mathrm {sccv}}) $  is $18$. 
% The percentage increase in 
%the number of xor gates used in source encoding therefore equals $\Delta_{\rm xor}=[(18-2)-(15-2)]/(15-2)
%= 23.1\%$. 
%Moreover, using \eqref{eq:Delta_latency} the percentage increase in the source encoding latency equals 
%$\Delta_{\rm latency, source}=12.5\%$.
%%\begin{equation}
%%\Delta_{\rm latency, source} = \frac{ (\lceil\log_2 (18-1)\rceil + \lceil\log_2 (15-1)\rceil ) - (\lceil\log_2 (15-1)\rceil + \lceil\log_2 (15-1)\rceil)}{\lceil\log_2 (15-1)\rceil + \lceil\log_2 (15-1)\rceil} =12.5\%.
%%\end{equation} 
%The percentage increase in 
%the number of LUTs used in C2V updates equals $(3*(18-2)-3*(15-2))/(3*(15-2))
%= 23.1\%$. 
%and the percentage increase in the decoding latency equals
%$\Delta_{\rm latency,dec}=10\%$.
%%{\color{black}
	%%\begin{equation}
	%%	\Delta_{\rm latency,dec}=\frac{ (\lceil\log_2 (18)\rceil + \lceil\log_2 (15)\rceil) - (2*\lceil\log_2 (15)\rceil)}{2*(\lceil\log_2 (15)\rceil-1) + 2*(\lceil\log_2 (4)\rceil-1)\rceil+\lceil\log_2 (5)\rceil-1}=10\%.
	%%\end{equation} }
	%}
%	

{\color{black} %\subsection{Example \#3}
	\noindent \underline{Example \#3:} We design a traditional DP-LDPC code based on $p_1=0.14$ %(entropy equals $0.58$ bit) 
	and then optimize the code using our proposed technique. 
	We reduce the source compression rate to adapt to the increased $p_1$. 
	We assume a source compression rate of $R_s=5/4$ and an overall symbol rate $R=1$. 
	Thus, the channel code rate is $R_c=R/R_s=4/5$. 
	We set $m_s$ and $n_s$ to $4$ and $5$, respectively. For a good channel code, $m_c$ should have a minimum value of $3$. Here, we set $m_c=3$. To match the required channel code rate, we set $n_c=7$ and $n_p=2$,
	i.e., there are 2 punctured VNs in the channel code. Using the differential evolution (DE) method in \cite{Chen2019}, we obtain 
	a traditional DP-LDPC code 
	${\textbf{B}}_{\mathrm J\_{\rm org}}^{0.14}={\textbf{B}}^{0.14}(x_i=0  \ \forall i)${\color{black}, where}  
	%\small
	\begin{equation}\label{B_0.14}
		{\textbf{B}}^{0.14}=\begin{array}{l}
			\left( {\begin{array}{ccccc|ccccccc}
					1 & 0 & 0 & 1 & 0 & 0 & 0 & 0 & 1 & 0 & 0 & 0   \\
					1 & 1 & 0 & 1 & 1 & 0 & 0 & 0 & x_1 & 1 & 0 & 0 \\
					0 & 1 & 1 & 1 & 1 & 0 & 0 & 0 & x_2 & x_4 & 1 & 0 \\
					0 & 0 & 1 & 0 & 1 & 0 & 0 & 0 & x_3 & x_5 & x_6 & 1 \\
					\hline
					0 & 0 & 0 & 0 & 0 & 1 & 0 & 0 & 1 & 0 & 1 & 1 \\
					0 & 0 & 0 & 0 & 0 & 0 & 1 & 0 & 2 & 0 & 1 & 1 \\
					0 & 0 & 0 & 0 & 0 & 0 & 1 & 3 & 1 & 2 & 0 & 1 \\
			\end{array}} \right) \\ 
		\end{array}.	
		% \tag {25}
	\end{equation}
	\normalsize
	%${\textbf{B}}^{0.14}$ is shown in \eqref{B_0.14}. 
	%This code is denoted by ${\textbf{B}}_{\mathrm J\_{\rm org}}^{0.14}$.
	Two VNs with the highest degrees are punctured, i.e., the $(n_s+4)$-th VN and the $(n_s+7)$-th VN in ${\textbf{B}}_{\mathrm J\_{\rm org}}^{0.14}$. The channel threshold of ${\textbf{B}}_{\mathrm J\_{\rm org}}^{0.14}$ is $-0.653$ dB while the Shannon limit is $-2.05$ dB at $p_1=0.14$.

	Next, we optimize ${\textbf{B}}_{\mathrm J\_{\rm org}}^{0.14}$ by replacing the identity matrix in the SCCV linking matrix with a lower or upper triangular base matrix. Here, we only consider the lower triangular base matrix to reduce 
	the searching complexity. Using the same DE method mentioned above, we obtain a new DP-LDPC code ${\textbf{B}}_{\mathrm J\_{\rm opt}}^{0.14} = {\textbf{B}}^{0.14}(x_1=x_3=1,x_2=x_4=x_5=x_6=0)$.
	%\eqref{B_0.14} when $x_1=x_3=1$ and $x_2=x_4=x_5=x_6=0$. This code is denoted as ${\textbf{B}}_{\mathrm J\_{\rm opt}}^{0.14}$. 
	Its channel threshold is $-0.840$ dB, which is $0.187$ dB lower than that of ${\textbf{B}}_{\mathrm J\_{\rm org}}^{0.14}$. 
	In Fig.~\ref{fig_added} ,
%			shows the SSER performance of ${\textbf{B}}_{\mathrm J\_{\rm opt}}^{0.14}$ and ${\textbf{B}}_{\mathrm J\_{\rm org}}^{0.14}$. We 
	we can see that ${\textbf{B}}_{\mathrm J\_{\rm opt}}^{0.14}$ outperforms ${\textbf{B}}_{\mathrm J\_{\rm org}}^{0.14}$ by about $0.2$~dB at {\color{black}an} SSER of $10^{-6}$.
	%
	%It can be shown that the percentage increases in source encoding complexity and latency 
	% are  both $33.3\%$, while there is no change in decoder complexity and decoding latency. 
}
\begin{figure}[htp]
{\color{black}
	\centering
	{\includegraphics[width=2.4in]{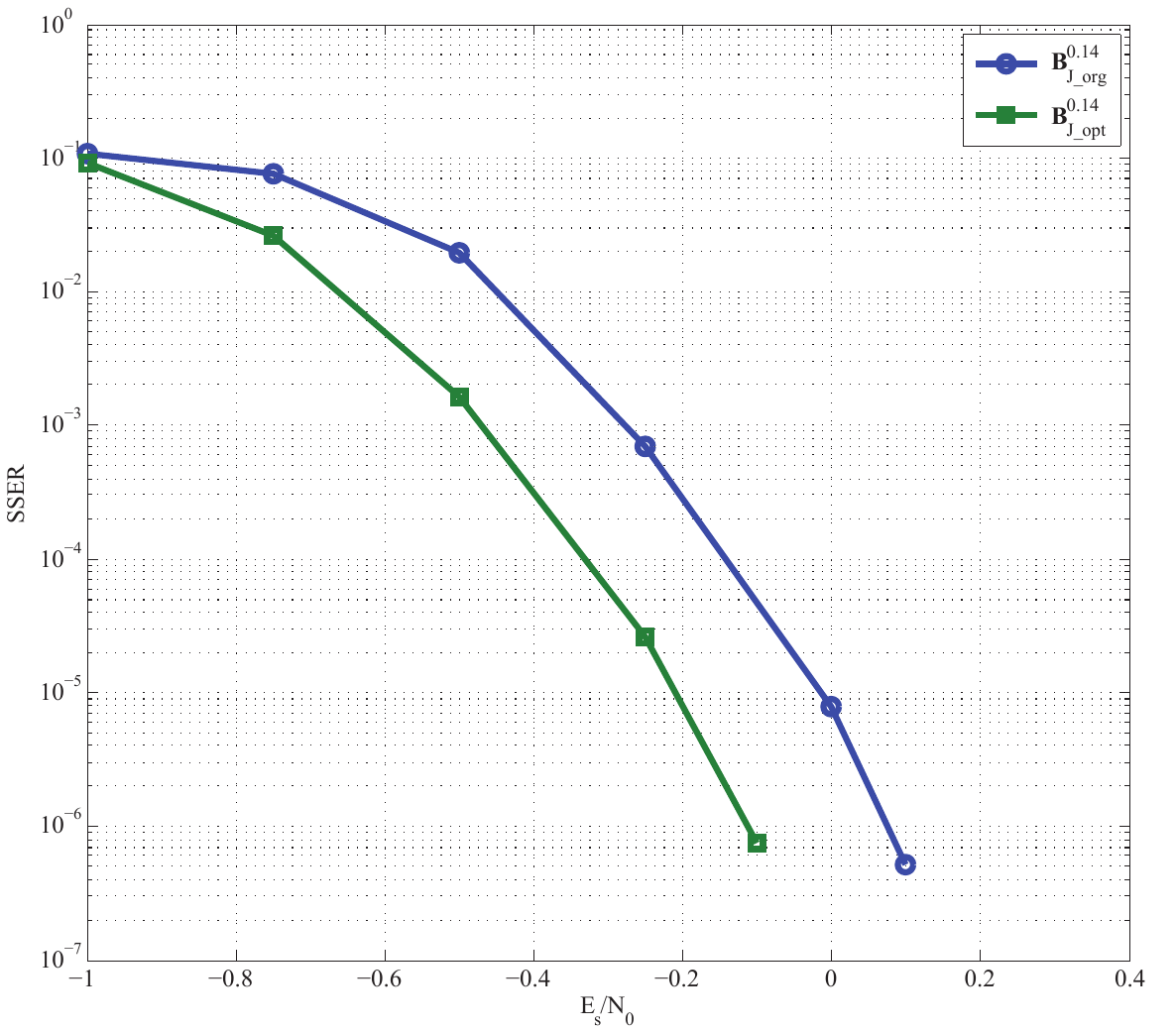}}
	\caption{{\color{black}SSER performance of ${\textbf{B}}_{\mathrm J\_{\rm org}}^{0.14}$ and ${\textbf{B}}_{\mathrm J\_{\rm opt}}^{0.14}$. $R=1$, $p_1=0.14$, and $N_s=z_1z_2n_s=4\times400\times5=8000$. {\color{black}The Shannon limit is $-2.05$ dB.}	
	}}
	\label{fig_added}}
\end{figure}

\begin{table*}[!t]
\centering
{\color{black}\caption{Comparison of complexity and latency between the new  DP-LDPC codes and the traditional DP-LDPC codes.}		
\scriptsize
	\begin{tabular}{|c|c|c|c|}
		\hline
		Code & ${\textbf{B}}_{\mathrm J\_{\rm opt1}}^{0.04}$ vs ${\textbf{B}}_{\mathrm J}^{0.04}$ 
		& $\textbf{B}_{\mathrm J4\_{\rm opt3}}^{0.01}$ vs $\textbf{B}_{\mathrm J4}^{0.01}$ & ${\textbf{B}}_{\mathrm J\_{\rm opt}}^{0.14}$ vs ${\textbf{B}}_{\mathrm J\_{\rm org}}^{0.14}$ \\
		\hline	
		Complexity increase in source encoding & $20\%$ & $23.1\%$ & $33.3\%$\\
		\hline
		$\Delta_{\rm latency,~source}$ & $0$  & $12.5\%$ & $33.3\%$\\
		\hline
		Complexity increase in decoding & $20\%$  & $23.1\%$ & $0$\\
		\hline
		$\Delta_{\rm latency,~dec}$ & $0$  & $10\%$ & $0$\\
		\hline
\end{tabular}}
\label{tab_latency}
\end{table*}

{\color{black}
\noindent \underline{Complexity and latency:}
Table~\ref{tab_latency} shows the complexity and latency increase of the new DP-LDPC codes compared with the 
traditional ones. Even though the percentage increase in the source encoding complexity
may seem large, the increase in the overall 
encoder complexity 
(when considering both source encoding and channel encoding) 
is much lower. 
}

{\color{black}\noindent \underline{Example \#4:} 
Finally, we compare our proposed code with the P-JSCCs in \cite{lau2021joint}, namely
AR3A-JSCC and AR4JA-JSCC, which are constructed at $p_1=0.04$ and $R=1$. 
To ensure the numbers of source symbols, punctured symbols, and transmitted symbols are
the same as those in \cite{lau2021joint}, we construct the following code 
%\small
\begin{equation}\label{BJopti_2}
	\begin{array}{l}
		\textbf{B}_{\mathrm J\_{\rm opt2}}^{0.04} %{\textbf{B}}_{\mathrm J}^{0.04}
		= \left( {\begin{array}{cccc|cccccc}
				2 & 2 & 1 & 1 & 0 & 0 & 0 & 0 & 1 & 1\\
				1 & 1 & 2 & 1 & 0 & 0 & 0 & 0 & 0 & 1\\
				\hline
				0 & 0 & 0 & 0 & 1 & 0 & 1 & 0 & 2 & 0\\
				0 & 0 & 0 & 0 & 0 & 1 & 0 & 1 & 1 & 1\\
				0 & 0 & 0 & 0 & 0 & 1 & 1 & 1 & 1 & 0\\
				0 & 0 & 0 & 0 & 0 & 0 & 1 & 0 & 2 & 0\\
		\end{array}} \right) \\ 
	\end{array},
\end{equation}
\normalsize
%$\textbf{B}_{\mathrm J\_{\rm opt2}}^{0.04}$ shown in \eqref{BJopti_2}, 
in which the last two VNs are punctured. $\textbf{B}_{\mathrm J\_{\rm opt2}}^{0.04}$  
is optimized based on $\textbf{B}_{\mathrm J\_{\rm opt1}}^{0.04}$ and the DE method. 
%
%For the SSER simulation results of these codes shown in \cite{lau2021joint}, the overall code length $N$ and the source length $N_s$ are set as $32000$ and $12800$, respectively. If we directly use $\textbf{B}_{\mathrm J\_{\rm opt1}}^{0.04}$ to compare them, its channel code length is $N_c=16000$ when we set the same source length as AR3A-JSCC and AR4JA-JSCC. In this case, $N\neq N_s+N_c$. To make a fair comparison between the new DP-LDPC and P-JSCC, we reconstruct the channel code in $\textbf{B}_{\mathrm J\_{\rm opt1}}^{0.04}$ and retain its source code and the upper triangular base matrix in $\textbf{B}'_{\mathrm {sccv}}$. The parameters of the channel code are $m_c=4$, $n_c=6$ and $n_p=2$. The channel code rate is still $0.5$. We use the DE method based on the JPEXIT algorithm to search entries in the channel code and obtain the code $\textbf{B}_{\mathrm J\_{\rm opt2}}^{0.04}$ shown in \eqref{BJopti_2}, whose ninth and tenth variable nodes are punctured.
The channel threshold of $\textbf{B}_{\mathrm J\_{\rm opt2}}^{0.04}$ is $-5.815$ dB, whereas those of AR4JA-JSCC and AR3A-JSCC are $-5.767$ dB and $-5.918$ dB, respectively. 
In Fig.~\ref{compAdd},
%		 plots the SSER performance of these three codes. 
%We 
we can see that ${\textbf{B}}_{\mathrm J\_{\rm opt2}}^{0.04}$ has a better SSER performance than AR4A-JSCC. AR3A-JSCC has the best waterfall performance among these codes but it suffers from an error floor at high $E_s/N_0$ region.		

\begin{figure}[htp]
	{\color{black}
		\centering
		{\includegraphics[width=2.3in]{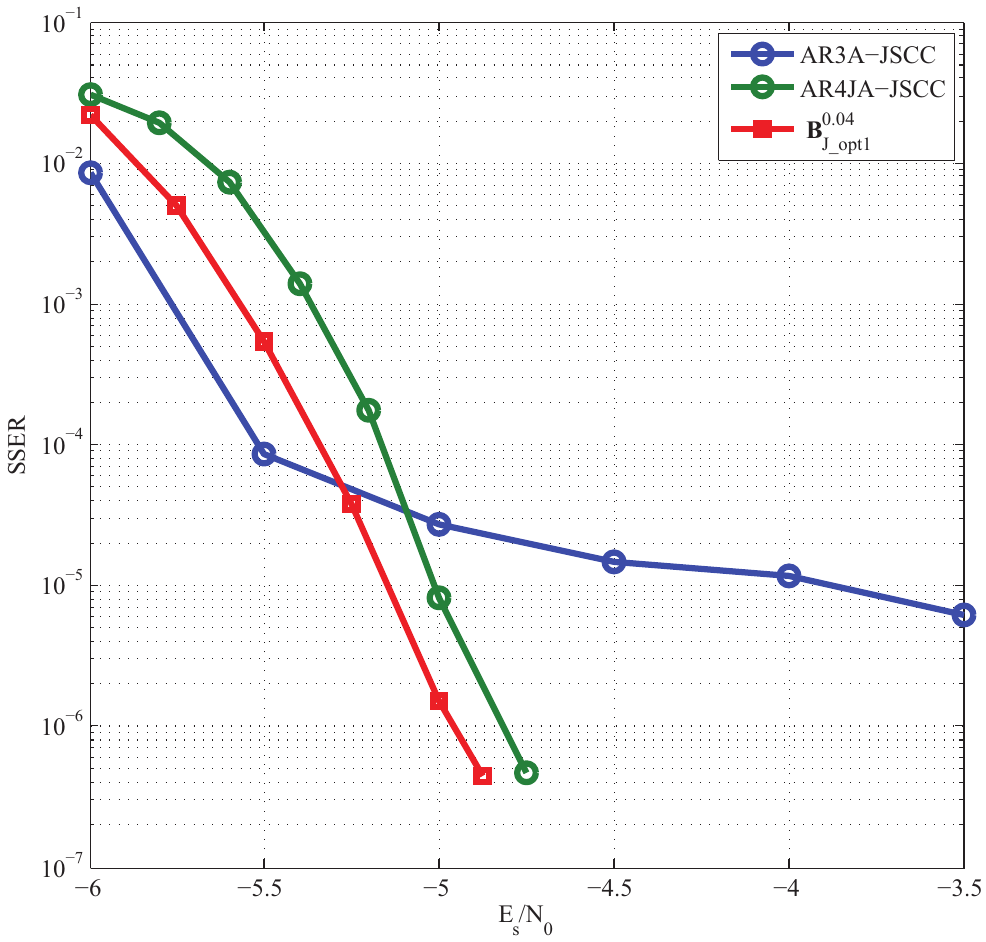}}
		\caption{SSER performance comparison of AR4JA-JSCC, AR3A-JSCC \cite{lau2021joint}, and ${\textbf{B}}_{\mathrm J\_{\rm opt1}}^{0.04}$. $R=1$, $p_1=0.04$, and $N_s=z_1z_2n_s=4\times800\times4=12800$. {\color{black}The Shannon limit is $-7.00$ dB.}} 	\label{compAdd}}
\end{figure}

\noindent \underline{Complexity and latency:} 
Since P-JSCCs do not have the specific structure (i.e., $\textbf{B}'_{\mathrm {sccv}}$ in \eqref{sccv_new}) as our constructed $\textbf{B}_{\mathrm J\_{\rm opt2}}^{0.04}$, the encoding complexities of AR4JA-JSCC and AR3A-JSCC are much higher than that of $\textbf{B}_{\mathrm J\_{\rm opt2}}^{0.04}$. 
Moreover, applying \eqref{eq:latency} to analyze AR4JA-JSCC, AR3A-JSCC, and $\textbf{B}_{\mathrm J\_{\rm opt2}}^{0.04}$
indicates that $\textbf{B}_{\mathrm J\_{\rm opt2}}^{0.04}$ achieves the lowest decoding latency.  
However, $\textbf{B}_{\mathrm J\_{\rm opt2}}^{0.04}$ has a larger ``highest row weight'' than 
AR4JA-JSCC and AR3A-JSCC, and hence a higher BP decoder complexity. 
Note that the decoder complexity for the three codes would become similar
if other decoding mechanisms, such as min-sum decoding, are used. 
%    
%	   For the new DP-LDPC codes, the complexity of obtaining the compressed source symbols is $\mathcal{O}(N_s+M_s)$. The complexity of channel encoding is $\mathcal{O}({N_c}^2)$ \cite{tanner2004ldpc}. Thus, the total encoding complexity for the new DP-LDPC codes is $\mathcal{O}(N_s+M_s)+\mathcal{O}({N_c}^2)$. For the P-JSCC, the encoding complexity is $\mathcal{O}({N}^2)\gg \mathcal{O}(N_s+M_s)+\mathcal{O}({N_c}^2)$ ($N=N_s+N_c$). Based on the analysis in Section \ref{sect:analysis}, we can know that, although $\textbf{B}_{\mathrm J\_{\rm opt2}}^{0.04}$ has higher decoding complexity than AR3A-JSCC and AR4JA-JSCC because its maximum row weight is larger, $\textbf{B}_{\mathrm J\_{\rm opt2}}^{0.04}$ has lower decoding latency than AR3A-JSCC and AR4JA-JSCC.
}

\section{Conclusion}\label{sect:conclusion}	
%In this paper, 
We have proposed a new class of DP-LDPC JSCC, which replaces
the identity matrix in the %source-check-channel-variable (SCCV) 
SCCV linking matrix of a traditional DP-LDPC JSCC with a lower or upper triangular matrix. 
Both theoretical and simulation results have demonstrated the
superiority of the proposed {\color{black}DP-LDPC JSCCs} over the traditional ones. 
\footnote{\color{black} We have investigated different node-puncturing combinations 
in the first three examples. We conclude 
that when optimizing the code design, only the entries corresponding to the punctured VN(s) may assume non-zero values. However, the results are not shown
due to space limitation.}

\bibliographystyle{IEEEtran}
\bibliography{BibTeX}

\end{document}